\RequirePackage{fix-cm}
\documentclass[onecollarge, natbib]{svjour2}
\bibpunct{[}{]}{,\!}{n}{}{,\!} 

\usepackage{amsmath}
\usepackage{amssymb}
\usepackage{mathtools}
\usepackage{bm}
\usepackage{color}
\usepackage{graphicx}
\usepackage{placeins}
\usepackage{supertabular}
\usepackage{xspace}
\usepackage{slashed}
\usepackage{hyperref}
\def\beq{\begin{equation}}
\def\eeq{\end{equation}}
\def\bea{\begin{eqnarray}}
\def\eea{\end{eqnarray}}

\def\br{{\bm r}}
\def\bk{{\bm k}}

\def\bp{{\bm p}}

\def\bB{{\bm B}}
\def\bE{{\bm E}}
\def\bP{{\bm P}}

\usepackage{mathptmx}      
\usepackage[mathscr,scaled=1.15]{urwchancal}
\DeclareFontFamily{OT1}{pzc}{}
\DeclareFontShape{OT1}{pzc}{m}{it}%
{<-> s * [1.15] pzcmi7t}{}
\DeclareMathAlphabet{\mathpzc}{OT1}{pzc}{m}{it}

\definecolor{purple}{rgb}{0.5,0,0.5}
\definecolor{blue}{rgb}{0.0,0,0.9}
\definecolor{prdblue}{rgb}{0.133,0.118,0.498}
\journalname{Few-Body Systems}
\begin{document}

\title{Femtoscopy of the origin of the nucleon mass }


\titlerunning{Femtoscopy of the proton mass}

\author{G.~Krein 
    \and
        T.~C.~Peixoto
}


\institute{G. Krein \at
            Instituto de F\'{\i}sica Te\'orica, Universidade Estadual Paulista, Rua Dr. Bento Teobaldo Ferraz, 271 - Bloco II, \\ 01140-070 S\~ao Paulo, SP, Brazil.\\[-2.5ex]
\and 
T. C. Peixoto  \at
            Instituto de F\'{\i}sica Te\'orica, Universidade Estadual Paulista 
            Rua Dr. Bento Teobaldo Ferraz, 271 - Bloco II, \\ 01140-070 S\~ao Paulo, SP, Brazil.\\[0.1ex]
            and \\[0.1ex]
            Instituto Federal de Educa\c{c}\~ao, Ci\^encia e Tecnologia de Sergipe,
            Rodovia Juscelino Kubitschek, s/n, \\ 49680-000 Nossa Senhora da Gl\'oria, SE, Brazil \\
            $\,$\\G. Krein [Corresponding author]\at
              \email{\href{mailto:gastao.krein@unesp.br}{gastao.krein@unesp.br}}\\
           }

\date{Received:
}
%
\maketitle
\begin{abstract}
We study the prospects of using femtoscopic low-momentum correlation 
measurements at the Large Hadron Collider to access properties of the
$J/\psi$-nucleon interaction. The QCD multipole expansion in terms
of the $J/\psi$ chromopolarizability relates the forward scattering
amplitude to a key matrix element to the origin of the nucleon mass 
problem, the average chromoelectric gluon distribution in the nucleon.
We use information on the $J/\psi$-nucleon interaction provided by 
lattice QCD simulations and phenomenological models to compute 
$J/\psi$-nucleon correlation functions. The computed correlation 
functions show clear sensitivity to the interaction, in particular 
to the $J/\psi$ chromopolarizability. 
\end{abstract}

%
\section{Introduction}
\label{intro}

What is the origin of the mass of protons and neutrons (nucleons) and, therefore, 
most of the universe's visible matter? Computer simulations of quantum chromodynamics 
(QCD) have given an answer to the question, namely: the mass comes mostly from the gluons 
and the nearly massless quarks. Yet, we are still unsatisfied and want more; 
we want to understand, to quote Wilczek~\cite{Wilczek:2008zz}: ``How did it happen?". 
With this mindset, an ever-growing effort is underway to find that kind of answer, both 
theoretically~\cite{Roberts:2016vyn,Roberts:2020udq} and experimentally~\cite{NAP25171:2008,Meziani:2020oks,
Chen:2020ijn}. The present work adds to this effort. We study the prospects of using 
femtoscopy in high-energy proton-proton and heavy-ion collisions for learning about the 
origin of the nucleon mass. 

Femtoscopy~\cite{Lednicky:1990pu} is a technique that makes it possible to obtain 
spatio-temporal information on particle production sources at the femtometer scale. 
Two-hadron momentum correlation functions carry such information~\cite{Heinz:1999rw,
Lisa:2005dd}. These correlation functions, remarkably, also carry information on 
low-energy hadron-hadron forces as final-state effects~\cite{Koonin:1977fh,Lednicky:1981su}. 
Relevant to the origin of the mass problem is the correlation function of a heavy 
quarkonium (such as $J/\psi$, $\eta_c$, $\Upsilon$,$\eta_b$) and a nucleon, for 
it gives direct access to the quarkonium-nucleon forward scattering amplitude. The QCD 
multipole expansion relates this amplitude to a key matrix element to the mass problem: 
the average chromoelectric gluon distribution in the nucleon~\cite{Peskin:1979va,Bhanot:1979vb,Kaidalov:1992hd,
Luke:1992tm,Kharzeev:1995ij,deTeramond:1997ny,Brodsky:1997gh,Ko:2000jx,Sibirtsev:2005ex,
Voloshin:2007dx,Krein:2017usp}. It is key to the problem because it relates to the trace of the QCD 
energy-momentum tensor in the nucleon, which defines the nucleon 
mass~\cite{SHIFMAN1978443,Donoghue:1987av}. 

The quarkonium-nucleon scattering amplitude is also accessible with $J/\psi$ and $\Upsilon$ 
electro- and photo-production experiments~\cite{Meziani:2020oks}. However, the kinematics
of the production process forbids direct access to the forward amplitude. In femtoscopy 
there are no such kinematics constraints. In addition, the two-particle correlation 
functions are measurable, in principle, down to zero relative momentum. Exemplar of 
femtoscopy's capabilities are the hyperon-proton and hyperon-hyperon correlation 
measurements in heavy-ion ($AA$), proton-ion ($pA$) and proton-proton ($pp$) collisions, 
ongoing for the last~15~years~\cite{Adams:2005ws,Agakishiev:2010qe,Adamczyk:2014vca,
Adamczewski-Musch:2016jlh,STAR:2018uho,Acharya:2018gyz,Acharya:2019kqn,Acharya:2019sms,
Acharya:2019yvb}. Closer to our interest in this paper is the recent feasibility
study~\cite{Chizali2019} of the $\phi-$proton system using data from $pp$ collisions 
collected by LHC's ALICE detector. In this and the envisioned quarkonium-nucleon case, 
the theoretical interpretation of the measurements profits from the absence of the 
Coulomb interaction and quantum-statistics, features that allow us to link a correlation
signal to a strong-interaction effect. Experimentally, however, the situation is not as 
clear-cut as in theory. Complications arise due to non-femtoscopic correlations, momentum 
resolution, and other experimental issues, which are always present in an experiment. 
Such effects need to be accounted for to extract the genuine strong-interaction correlation. 
In addition, knowledge of the particle source emission form and size are important 
issues for the theoretical interpretation of the data. Notwithstanding experimental 
issues, the subject's relevance and successes with the hyperon-proton system and 
preliminary results on the $\phi$-proton system motivate our prospective study. 
We hope our results will motivate experimental studies as well.

We focus on the $J/\psi$-nucleon ($J/\psi\,N$) system, having in
mind experiments at the LHC. The motivation for preferring $J/\psi$ 
over the other heavy quarkonia is twofold. First, $J/\psi$ production yields 
and weak decay rates are relatively high{\textemdash}the branching ratios of 
decays to $e^+e^-$ and $\mu^+\mu^-$ is of the order of 6\% each. Second, there 
is some theoretical knowledge on the $J/\psi\,N$ system from lattice QCD
studies~\cite{Yokokawa:2006td,Liu:2008rza,Kawanai:2010ru,Kawanai:2010ev,
Kawanai:2011zz,Alberti:2016dru,Sugiura:2017vks,Skerbis:2018lew,Sugiura:2019pye}.
These studies revealed an attractive and not very strong interaction. One 
lattice study~\cite{Beane:2014sda} found a bound $\eta_c\,N$ state  
with $20$~MeV binding, a binding energy much larger than phenomenological 
expectations~\cite{Krein:2017usp,Hosaka:2016ypm}. For our femtoscopic study, we use the
lattice results of Refs.~\cite{Yokokawa:2006td,Liu:2008rza,Kawanai:2010ru},
which provide values for the $S-$wave scattering length $a_0$ and effective range 
$r_0$ parameters, and of Ref.~\cite{Kawanai:2010ev} which gives in addition an 
$S-$wave $J/\psi\,N$ potential extracted with the HALQCD method. These lattice 
results were obtained either with quenched gluon configurations~\cite{Yokokawa:2006td,
Kawanai:2010ru} or large pion masses~\cite{Liu:2008rza,Kawanai:2010ev} and,
therefore, require extrapolation to the physical mass. A~recent 
study~\cite{TarrusCastella:2018php} performed such an extrapolation with an 
effective field theory (EFT) specially set 
up for studying the low-energy the quarkonium-nucleon interaction. The EFT, dubbed 
QNEFT, obtained expressions for $a_0$ and $r_0$ at leading order (LO) and 
next-to-leading order (NLO) in the pion mass. The extrapolated values preserve 
the overall qualitative picture of a weakly attractive interaction. 
The QNEFT also predicts a model-independent van der Waals type of potential 
of range~$1/2m_\pi \simeq 0.7$~fm, with a strength controlled by the $J/\psi$
chromopolarizability. We note that the $J/\psi\,N$ system can have 
spin~1/2 or spin~3/2. The early latttice results of Refs.~\cite{Yokokawa:2006td,
Liu:2008rza,Kawanai:2010ru,Kawanai:2010ev} found no significant hyperfine
splitting in $a_0$ and $r_0$. Therefore, the QNEFT extrapolations should be
considered as spin-1/2 and spin-3/2 averages. Interestingly, a recent
lattice study~\cite{Sugiura:2017vks} within the HALQCD method, but with 
larger lattice volumes and a relativistic heavy quark action for the charm 
quark, found a somewhat sizable hyperfine splitting, $a^{1/2}_0 \simeq 1.7 \, a^{3/2}_0$. 

Further knowledge on the $J/\psi\,N$ interaction comes from phenomenological models. 
We consider two models, those of Refs.~\cite{Anwar:2018bpu,Eides:2017xnt}. 
Both models address the interaction of a charmonium with a nucleon within the 
hadro-charmonium picture~\cite{Dubynskiy:2008mq}. In this picture, a 
charmonium interacts as a compact object within the volume of a light hadron. The
model of Ref.~\cite{Anwar:2018bpu} treats the nucleon as a spherical finite well.
This is a very simple but insightful model, as it can be solved analytically. The model
of Ref.~\cite{Eides:2017xnt} uses the chiral soliton quark model ($\chi$QSM) of 
Ref.~\cite{Goeke:2007fp}. The model uses the QCD trace anomaly~\cite{Chanowitz:1972vd,
Crewther:1972kn,Chanowitz:1972da,Freedman:1974gs,Collins:1976yq} to obtain an effective
$J/\psi\,N$ potential in terms of the gluon energy-momentum density inside a nucleon
calculated with the $\chi$QSM. 

We use the knowledge on the $J/\psi\,N$ system from lattice QCD simulations,
extrapolated to the physical pion mass with the QNEFT~\cite{TarrusCastella:2018php},
and from the phenomenological models of Refs.~\cite{Anwar:2018bpu,Eides:2017xnt} 
to make predictions for the $J/\psi\,N$ correlation function. In the next section 
we review the basics of femtoscopy. We discuss limits for which the correlation 
function can be linked directly to the scattering amplitude and how this allows us 
to link the correlation function to the average chromoelectric gluon distribution 
in the nucleon. In Section~\ref{sec:res} we present predictions for the $J/\psi\,N$ 
correlation function. We study the dependence of the correlation function on source 
emission size and on parameters of the interaction. Section~\ref{sec:concl} presents 
a summary of our work.

%
\section{Correlation function and scattering amplitude}
\label{sec:correl}

The observable of interest in femtoscopy is a two-hadron correlation function 
$C(\bp_1,\bp_2)$ of measured hadron momenta $\bp_1$ and
$\bp_2$~\cite{Heinz:1999rw,Lisa:2005dd}. The extraction of the experimental 
correlation function involves computing the ratio of two yields, $A(k)/B(k)$, 
where $k=|\bk|$ with $\bk = {\bp}_1 = - {\bp}_2$, the relative momentum  
in the center of mass of the pair\footnote{The total momentum is
$\bP = \bp_1 + \bp_2$ and the relative momentum is 
$\bk = (m_2 \bp_1 - m_1 \bp_2)/(m_1 + m_2)$, where $m_1, m_2$ are the particles' 
masses. In the center of mass frame, $\bP = 0$ and $\bp_1 = -\bp_2$; hence, 
the relative momentum is $\bk = \bp_1 = - \bp_2$.}. $A(k)$ is the coincidence yield (or signal 
distribution), formed by pairs with a given $k$ produced in the same 
collision event, and $B(k)$ is the uncorrelated yield (or background 
distribution), formed by pairs with the same~$k$ but collected from different 
collision events. Corrections for non-femtoscopic correlations in $A(k)$ not 
accounted for in $B(k)$, and other experimental effects, are taken into account 
by a multiplicative factor~$\xi(k)$, so that $C(k) = \xi(k) \times A(k)/B(k)$. 
The theoretical interpretation of the experimental correlation function is 
usually based on the Koonin-Pratt formula~\cite{Koonin:1977fh,Pratt:1984su}:
\begin{equation}
C(k) = \xi(k) \, \frac{A(k)}{B(k)} 
= \int d^3\!r \; S_{12}(\br)\, |\psi(\bk,\br)|^2 \ .
\label{KP}
\end{equation}
Here $\psi(\bk,\br)$ is the relative wave function of the pair and $S_{12}(\br)$ 
a static emission source, a pair's relative distance distribution in 
the pair frame. Refs.~\cite{Heinz:1999rw,Lisa:2005dd,Bauer:1993wq,Anchishkin:1997tb}
address in depth the validity of the assumptions and approximations behind Eq.~(\ref{KP}).

The current knowledge of the quarkonium-nucleon interaction lets us assume 
that it affects only the wave function's $S-$wave component. Therefore, 
separating from $\psi(\bk,\br)$ the $l=0$ component, $\psi_0(k,r)$,
which contains the effects of the strong interaction, we can write $\psi(\bk,\br)$
as
\begin{equation}
\psi(\bk,\br) = e^{i \bk\cdot\br} + \psi_0(k,r) - j_0(kr) ,  
\label{Eq:wfa}
\end{equation}
where $j_0(kr)$ is the $l=0$ spherical Bessel function, the $S-$wave component of the 
non-in\-ter\-acting wave function. Taking a one-parameter Gaussian form for the
emission source, $S_{12}(r) = (4\pi R^2)^{-3/2} \exp(-r^2/4R^2)$, the Koonin-Pratt 
formula can be written as 
\begin{equation}
\hspace{-0.10cm}
C(k) = 1 + \frac{4\pi}{(4\pi R^2)^{3/2}} \int^\infty_0 dr \, r^2 \, e^{-r^2/4R^2}
\left[|\psi_0(k,r)|^2
- |j_0(kr)|^2 \right] .
\label{Ck}
\end{equation}
The Gaussian form is a common choice as it simplifies the analysis~\cite{Ohnishi:2016elb,
Morita:2016auo,Haidenbauer:2018jvl,Haidenbauer:2020kwo}, but it represents an  
experimental issue, as we discuss in the next section. 

Two length scales in $C(k)$ are important for extracting properties of the interaction: 
the emission source radius~$R$ and the two-particle interaction effective range. 
When these lengths are of comparable size,  most of the emitted particles are under 
the influence of the interaction. Therefore, one needs the pair wave function 
$\psi_0(k,r)$ in the entire range $0 \leq r \leq \infty$ of integration in Eq.~(\ref{Ck}). 
One can use either the Schr\"odinger equation or the Lippmann-Schwinger equation to obtain 
$\psi_0(k,r)$; the latter is well suited for treating nonlocal potentials and coupled 
channels~\cite{Haidenbauer:2018jvl,Haidenbauer:2020kwo}. We deal with local $J/\psi\,N$ 
potentials in this paper and use the Schr\"odinger equation to obtain $\psi_0(k,r)$.

On the other hand, for $R$ much larger than the effective range of the interaction, 
most of the emitted pairs are not under the influence of the interaction. Then, one
can replace $\psi_0(k,r)$ with its asymptotic form
\beq 
\psi^{asy}_0(k,r) = \frac{\sin (kr+\delta_0)}{kr} 
=  e^{-i \delta_0} \left[j_0(kr) + f_0(k) \frac{e^{ikr}}{r}\right] 
\hspace{0.5cm}\text{with}\hspace{0.5cm}
f_0(k) = \frac{e^{i\delta_0} \sin \delta_0}{k},
\label{psi-asy}
\eeq 
where $f_0(k)$ is the scattering amplitude and $\delta_0$ the phase shift. 
However, use of $\psi^{asy}_0(k,r)$ in place of $\psi_0(k,r)$ in 
Eq.~(\ref{Ck}) for all values of~$r$ incurs in error for pairs emitted from 
within the range of the interaction. One way to account for this 
error uses~\cite{Lednicky:1981su} effective range theory~\cite{Bethe:1949yr} 
to evaluate the correction $|\psi_0(k,r) - \psi^{\rm asy}(k,r)|^2$ for 
$r \simeq 0$. Using $\psi^{asy}_0(k,r)$ in place of $\psi_0(k,r)$ 
in Eq.~(\ref{Ck}) and including the short-range correction, one obtains 
for $C(k)$~\cite{Lednicky:1981su}: 
\begin{eqnarray}
C(k) &=& 1 + \frac{|f_0(k)|^2}{2R^2} \left(1 - \frac{r_0}{2 \sqrt{\pi} R}\right) 
+ \frac{2 {\rm Re} f_0(k)}{\sqrt{\pi} R} F_1(2 kR)
- \frac{{\rm Im} f_0(k)}{R} F_2 (2 kR),
\label{Ck-f0}
\end{eqnarray}
where $r_0$ the effective range parameter that appears in the 
effective range expansion (ERE) of $f_0(k)$:
\beq 
f_0(k) = \frac{1}{k \cot \delta_0 - ik} \overset{k\approx 0}{\longrightarrow} 
\frac{1}{-\frac{1}{a_0} + \frac{1}{2} r_0\,k^ 2 - ik} ,
\label{efr}
\eeq 
and $a_0$ is the scattering length, and
\beq 
F_1(x) = \frac{1}{x} \int^x_0 dt \, e^{\, t-x}, \hspace{1.0cm} 
F_2(x) = \frac{1}{x}\left(1 - e^{-x^2}\right).
\eeq 
The short-range correction is the term ${r_0}/{2 \sqrt{\pi} R}$ 
in Eq.~(\ref{Ck-f0}); taking $r_0=0$ in that equation amounts to replacing 
$\psi_0(k,r)$ by $\psi^{asy}_0(k,r)$ everywhere in Eq.~(\ref{Ck}). Of course, 
to make sense, the correction must be small. Eq.~(\ref{Ck-f0}) is known as the 
Lednicky-Lyuboshits (LL) formula. For small values of $k$, one can use the
ERE for $f_0(k)$ and then Eq. (5) becomes universal, in the sense that it
depends only on $a_0$, $r_0$, and $R$; no further knowledge is required 
to compute the correlation function. 

If the LL formula is applicable, $C(k)$ at small~$k$ gives direct access to 
the matrix element of the average chromoelectric gluon distribution in the 
nucleon, $\langle N| (g\bE^a) \cdot (g\bE^a)|N\rangle \equiv 
\langle (g\bE)^2 \rangle_N$, where $\bE^a, a=1,\cdots,8,$ is the 
chromoelectric gluon field, $g$ the strong coupling constant, and $|N\rangle$ 
the nucleon state. The average $\langle (g\bE)^2 \rangle_N$ appears in 
the low energy $J/\psi\,N$ forward scattering amplitude within the QCD multipole 
expansion framework, in that the heavy quarkonium behaves like a small color 
dipole interacting with soft gluon fields of the nucleon~\cite{Peskin:1979va,
Bhanot:1979vb,Kharzeev:1995ij,Sibirtsev:2005ex,Voloshin:2007dx,Krein:2017usp}. 
For an $S-$wave dominating interaction, as it seems to be the case for $J/\psi\,N$, 
the forward amplitude at small~$k$ is real and given by $f_{\rm forw}(k) = 
f_0(k) \simeq - a_0$~\cite{Kaidalov:1992hd,Sibirtsev:2005ex,Voloshin:2007dx}, with 
\beq
a_0 = \frac{\mu}{4\pi} \, \alpha_{J/\psi} \, \langle (g\bE)^2 \rangle_N,
\label{f0-a0}
\eeq
where $\mu$ is the $J/\psi\,N$ reduced mass, $\alpha_{J/\psi}$ the $J/\psi$ 
chromopolarizability, with $\langle (g\bE)^2 \rangle_N$ evaluated with the
nucleon at rest. Therefore, if the value of $\alpha_{J/\psi}$ is known, 
one can obtain $\langle (g\bE)^2 \rangle_N$ from the $a_0$ extracted from
the measured $C(k)$ via the LL formula. Unfortunately, at present 
$\alpha_{J/\psi}$ is not well constrained by data; as such, while the 
situation persists, there will be an associated uncertainty in the extraction 
of $\langle (g\bE)^2 \rangle_N$. We note that this difficulty is not restricted 
to femtoscopy; any experiment attempting to extract $\langle (g\bE)^2 \rangle_N$ 
from the scattering length via Eq. (8), as for example $J/\psi$ electro- and 
photo-production experiments~\cite{Meziani:2020oks}, share with femtoscopy 
this difficulty. 

If the LL formula is not applicable, the link between the measured 
$C(k)$ and $\langle (g\bE)^2 \rangle_N$ is more indirect, as one needs
a model for the interaction to extract the scattering length from the 
experimental $C(k)$. The $\Sigma^+\Sigma^+$ interaction, discussed in 
Ref.~\cite{Haidenbauer:2018jvl}, is an example indicating failure of 
the LL formula (cf. Fig.~[5] of that reference), the failure being
associated with the large value of the effective range parameter~$r_0$.
In the next section, we show results for $C(k)$ computed with two models 
for the $J/\psi\,N$ interaction for which, depending on the value of 
$\alpha_{J/\psi}$, $r_0$ is large.    

To conclude this section, we note that the connection between $\langle (g\bE)^2 \rangle_N$ 
and the trace of the QCD energy-momentum tensor and the nucleon mass comes through the 
inequality~\cite{Voloshin:2007dx}: 
\bea
\langle N|\left[ \left(g\bE^a\right)^2 - \left(g{\bB^a}\right)^2 \right] |N\rangle 
= - \frac{1}{2} \langle N|g^2 G^a_{\mu\nu} G^{a \mu \nu}|N\rangle
= \frac{16\pi^2}{9} m_N \leq \langle (g\bE)^2 \rangle,
\label{inequ}
\eea
where ${\bB^a}$ is the chromomagnetic field. Here we used the trace-anomaly relationship~\cite{Chanowitz:1972vd,Crewther:1972kn,
Chanowitz:1972da,Freedman:1974gs,Collins:1976yq}
\beq
T^\mu_\mu(x) = - \dfrac{9}{32\pi^2} \, g^2 G^a_{\mu\nu}(x) G^{a \mu \nu}(x),  
\label{Tmunu}
\eeq 
valid in the chiral limit. The last inequality in Eq.~(\ref{inequ}) follows 
from the fact that $\langle N|\left(g{\bB^a}\right)^2|N\rangle \ge 0$. The normalization
of the nucleon state we use is such that the expectation value of $T^{00}$ 
is the energy~\cite{Donoghue:1987av,Krein:2017usp}. Away from the chiral limit, 
Eq.~(\ref{Tmunu}) contains the contribution from the quark-mass term of the QCD 
Lagrangian, whose contribution to $m_N$ seems to be small~\cite{Krein:2017usp}.

%
\section{Numerical results and discussions}
\label{sec:res} 

Here we present predictions for the $J/\psi \,N$ correlation function. We use 
information on the $J/\psi\,N$ interaction from lattice QCD simulations,
extrapolated to the physical pion mass with QNEFT~\cite{TarrusCastella:2018php},
and from the phenomenological models of Refs.~\cite{Anwar:2018bpu,
Eides:2017xnt}. Both QNEFT and the phenomenological models consider spin-1/2 and 
spin-3/2 degeneracy. The radius~$R$ of the assumed one-parameter 
Gaussian form for the emission source is treated as a free parameter; we present
results for $R = 1$~fm and $R = 3$~fm. The lower value of $R$ is thought appropriate 
for $pp$ collisions, and the highest for $pA$ and $AA$ collisions. These are smaller
source sizes as those employed in hyperon-nucleon studies~\cite{Haidenbauer:2018jvl,
Chizali2019,Haidenbauer:2020kwo}. The use of smaller source sizes is motivated by 
evidence that emission sources scale with the inverse of the emitted particles 
masses~\cite{Lisa:2005dd}. We recall that the source in femtoscopy refers to the 
``region of homogeneity'', the region from which particle pairs with a certain 
momentum are most likely emitted, which is significantly smaller than the size 
of the entire source emitting particles~\cite{Heinz:1999rw,Lisa:2005dd}. 

We begin presenting results using the QNEFT-extrapolated lattice 
information~\cite{TarrusCastella:2018php} for the $J/\psi\,N$
interaction. The interaction contains contact terms and a long-range,
model independent van der Waals potential:
\beq 
V_{\rm vdW}(r) = \frac{3 g^2_A}{128\pi^2 F^2} \left\{
c_{di}\left[6 + m_\pi r(2+m_\pi r)(6+m_\pi r (2+m_\pi r))\right] 
+  c_m m^2_\pi r^2 (1 + m_\pi r)^2 \right\}\,
\frac{e^{-2m_\pi r}}{r^6},
\label{VvdW}
\eeq 
where $g_A = 1.27$ is the nucleon axial charge, $F = 93$~MeV the pion decay constant, 
and $c_{di}$ and $c_m$ are low-energy constants; $c_{di}$ reflects the hadronization 
into two pions of the soft-gluon coupling $(E^a)^2$ mediating the interaction between
the nucleon and $J/\psi$, and $c_m$ relates to the light-quark masses. The value of
$c_{di}$ can be determined by using the QCD trace anomaly~\cite{Brambilla:2015rqa}. 
Explicit expressions for $c_{di}$ and $c_m$ are given in Eq.~(5) of Ref.~\cite{TarrusCastella:2018php}; 
they also depend on the $J/\psi$ chromopolarizability, $\alpha_{J/\psi}$ (denoted by $\beta$ 
in that reference), the only parameter in Eq.~(\ref{VvdW}) not well constrained 
by experiment. Ref.~\cite{TarrusCastella:2018php} extracted the value 
$\alpha_{J/\psi} = 0.24~{\rm GeV}^{-3}$ by fitting $V_{\rm vdW}(r)$ to the 
$J/\psi\,N$ HALQCD potential~\cite{Kawanai:2010ev}. 
Given the uncertainties in the lattice values for $a_0$ and $r_0$~\cite{Yokokawa:2006td,
Liu:2008rza,Kawanai:2010ru}, the corresponding QNEFT-extrapolated values are
$-0.71~{\rm fm} \leq a_0 \leq -0.35~{\rm fm}$, and $1.29~{\rm fm} \le r_0 
\leq 1.35~{\rm fm}$. We note that a much smaller value for the scattering length,
$a_0 = -0.05$~fm, was obtained in Ref.~\cite{TarrusCastella:2018php} when 
identifying the LO term of the QNEFT scattering amplitude with the one
in Ref.~\cite{Sibirtsev:2005ex}, which uses the multipole expansion for 
quarkonium interaction. A low value for $a_0$ was also obtained in an earlier 
study~\cite{Hayashigaki:1998ey} using QCD sum rules, $a_0 = - 0.10$~fm. 
Smaller values were also extracted from old~\cite{Gryniuk:2016mpk} and 
recent~\cite{Strakovsky:2019bev,Pentchev:2020kao} photoproduction 
data{\textemdash}it should be noted that the extraction of the scattering length 
in such experiments rely on assumptions and extrapolations to zero momentum 
transfer, as kinematics forbids direct access to the forward amplitude. 

\begin{figure}[!htbp]
\centerline{ 
\includegraphics[scale=0.825]{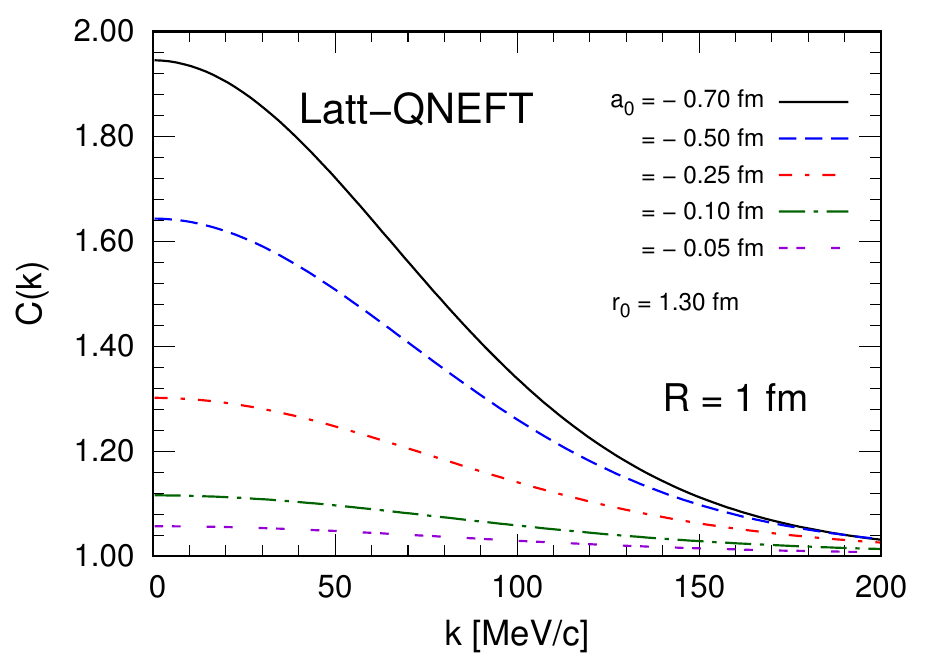}\hspace{-0.1cm}
\includegraphics[scale=0.825]{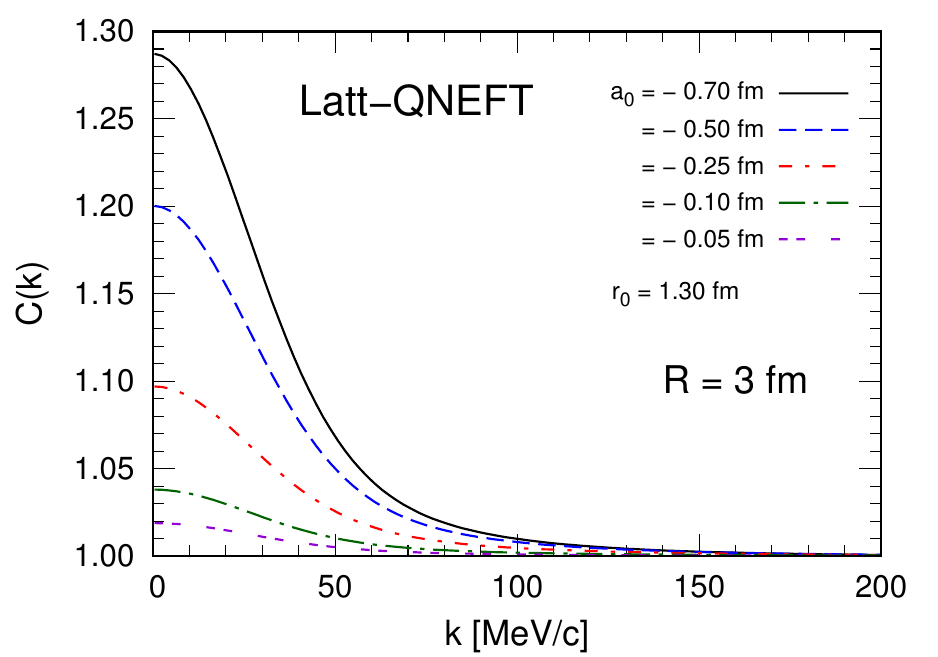}
}
\caption{$J/\psi\,N$ correlation function computed with Eq.~(\ref{Ck-f0})
for two values of the source radius~$R$. The $a_0$ and $r_0$ values are from 
lattice QCD simulations~\cite{Yokokawa:2006td,Liu:2008rza,Kawanai:2010ru,
Kawanai:2010ev} extrapolated to the physical pion mass with 
QNEFT~\cite{TarrusCastella:2018php}.  
}
\label{Fig1}
\end{figure}

The values of range of the van der Waals force, $1/2m_\pi$, strength at
$r= 1/2m_\pi$, 3~MeV, and the values of $a_0$ and $r_0$, justify use of 
Eq.~(\ref{Ck-f0}) to compute $C(k)$ for $R=1$ and $R=3$. Since the value 
of the effective range $r_0$ does not vary much within the uncertainties, 
we fix it to $r_0 = 1.3$~fm and present results for the scattering length 
varying in the range $ - 0.7~{\rm fm} \leq a_0 \leq - 0.05~{\rm fm}$. 
Fig.~\ref{Fig1} displays results for $C(k)$ for two values of the source 
radius,~$R = 1$~fm and $R = 3$~fm.  

Figure~\ref{Fig1} reveals the expected trend about correlation strength 
and source size, in that stronger correlations happen for smaller sources. 
From this perspective alone, $pp$ collisions are preferred over collisions 
with heavy ions, although production yields are higher in the latter. 
Regarding the predictions of the correlation strengths for the 
different values of the scattering length~$a_0$, they are small for 
$R=3$~fm for the smallest values of $a_0$; namely $C(0) \simeq 1.04$ for 
$a_0 = - 0.10$~fm, $C(0) \simeq 1.02$ for $a_0 = - 0.05$~fm. On the
other hand, for $R=1$, the correlation strengths are comparable
to those extracted for the $\phi\,N$ system in Ref.~\cite{Chizali2019}.   

Next, we present results for $C(k)$ computed with scattering wave functions 
obtained with the Schr\"odinger equation for the potentials of 
Refs.~\cite{Anwar:2018bpu} and \cite{Eides:2017xnt}. The first uses
a spherical finite well of radius $R_N$, given in Eq.~(4) of 
Ref.~\cite{Anwar:2018bpu}:
\begin{equation}
V(r) = \left\{ \begin{array}{ccc}
- \frac{2\pi}{3}\left( \frac{\alpha_{J/\psi}}{R^3_N}\right) m_N 
& \mbox{for} & r < R_N \\[0.3true cm]
0   & \mbox{for} & r > R_N
\end{array}  \right.  \mbox{ }.
\label{sphw}
\end{equation}
The second potential is Eq.~(11) of Ref.~\cite{Eides:2017xnt}:
\bea
V(r) = - \alpha_{J/\psi} \frac{4\pi^2}{b} \left(\frac{g^2}{g^2_s}\right)
\left[ \nu \rho_E(r) - 3 p(r) \right],
\label{eides}
\eea 
with $b = (11 N_c - 3 N_f)/3 = 27/3$ and $g^2/g^2_s = 1$~(adequate values for $J/\psi$) and 
$\nu = 1.5$. The energy density $\rho_E(r)$ and pressure $p(r)$, given in
terms of matrix elements of $T^{00}(x)$ and $T^\mu_\mu(x)$ in the nucleon state, 
are computed with the $\chi$QSM in Ref.~\cite{Goeke:2007fp}{\textemdash}we scanned 
their profiles from Fig.~1 of this reference. In both potentials, only 
$\alpha_{J/\psi}$ is not well constrained by theory and experiment.
Therefore, we present results for $C(k)$ for four values of $\alpha_{J/\psi}$, covering 
a wide range of values commonly used in the literature, namely: 
$\alpha_{J/\psi} = 2~{\rm GeV}^{-3}$~\cite{Sibirtsev:2005ex,Voloshin:2007dx},
$1.60~{\rm GeV}^{-3}$~\cite{Polyakov:2018aey}, $0.54~{\rm GeV}^{-3}$~\cite{Hayashigaki:1998ey},
$0.24~{\rm GeV}^{-3}$~\cite{TarrusCastella:2018php}. We set the well radius to $R_N = 1$~fm, as the 
results for $C(k)$ with $R=3$~fm follow the trend shown in Fig.~\ref{Fig1}. 

Table~\ref{tab:a0r0} lists the scattering length and effective range parameters
extracted from the wave functions. One sees that the $a_0$ values corresponding to the 
different $\alpha_{J/\psi}$, for both potentials, are in close correspondence to 
those used in Fig.~\ref{Fig1}. Table~\ref{tab:a0r0} also reveals the well known 
fact that when $|a_0|$ is much smaller than the actual range of the potential 
$R_{\rm range}$, $r_0$ can be very different from~$R_{\rm range}$. In this case,
use of the ERE for $f_0$ in the Lednicky model becomes 
problematic~\cite{Haidenbauer:2018jvl}.

\begin{table}[!htb]
\caption{Scattering length and effective range parameters (in fm) corresponding
to the finite well with $R_N=1$~fm and $\chi$SQM potentials, Eqs.~(\ref{sphw}) and (\ref{eides}),
for several values of $\alpha_{J/\psi}$ (in ${\rm GeV}^{-3}$). 
}
\label{tab:a0r0}       
\begin{center}
\begin{tabular}{lcccc}\hline
\\[-0.10true cm]
& \multicolumn{2}{c}{Finite well}  & \multicolumn{2}{c}{$\chi$SQM} 
\\
\\[-0.2true cm]
\cline{2-3}\cline{4-5}
\\[-0.2true cm]
$\alpha_{J/\psi}$  & $a_0$    & $r_0$     & $a_0$     & $r_0$     \\[0.2true cm] \hline \\
$2.00$             & $-0.68$  & $1.59$    & $-0.42$   & $1.86$     \\ \\[-0.2true cm]
$1.60$             & $-0.47$  & $1.86$    & $-0.30$   & $2.25$     \\[0.25true cm]
$0.54$             & $-0.12$  & $4.50$    & $-0.08$   & $6.00$     \\[0.25true cm]
$0.24$             & $-0.05$  & $9.46$    & $-0.03$   & $13.05$    \\[0.10true cm]
\hline
\end{tabular}
\end{center}
\end{table}

\begin{figure}[!htbp]
\centerline{ 
\includegraphics[scale=0.825]{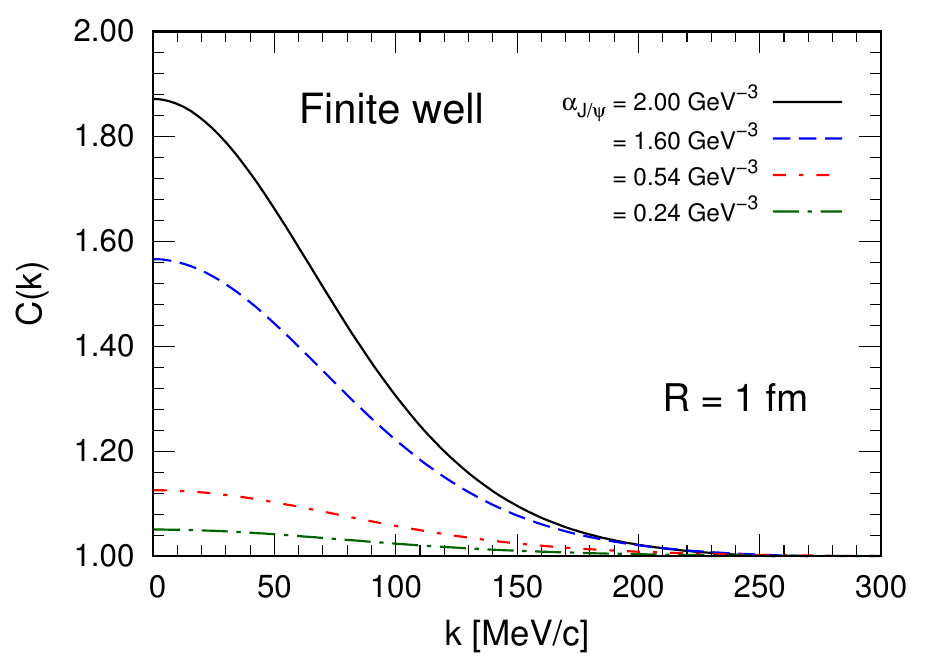}\hspace{-0.1cm}
\includegraphics[scale=0.825]{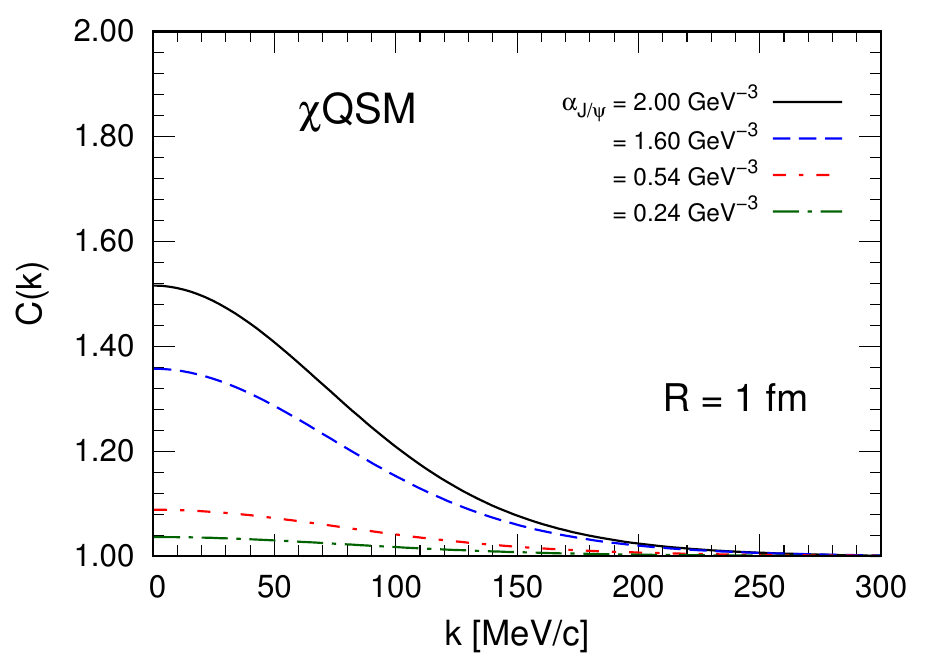}
}
\caption{$J/\psi\,N$ correlation function calculated with pair 
wave functions $\psi_0(k,r)$ computed with the Schr\"odinger equation
for the entire range $0 \leq r \leq \infty$ of the integrand of Eq.~(\ref{Ck}). 
The $\psi_0(k,r)$ correspond to the potentials in Eqs.~(\ref{sphw}) 
and (\ref{eides}).
}
\label{Fig2}
\end{figure}

Figure~\ref{Fig2} displays the results for $C(k)$ using the wave functions 
corresponding to the two potentials. The results for the finite well
are very similar to those in Fig.~\ref{Fig1}. This is not surprising given 
the fact already mentioned regarding the extracted values of $a_0$. 
The $C(k)$ values for the $\chi$SQM are a little smaller than those for 
the finite well because the $a_0$ values are a little smaller for the former. 

Given the results in Figs.~1~and~2, one sees that if the $J/\psi\,N$ interaction 
turns out to be very weak, to extract information from femtoscopic measurements 
will be very challenging. As already mentioned, there will always be experimental 
errors related to source size, momentum resolution, etc that will not allow to 
resolve weak correlations. In concrete terms, there seems to be room for 
optimism if the value of the chromopolarizability $\alpha_{J/\psi}$ turns out 
to be above $0.24~{\rm GeV}^{-3}$.   

To conclude, we note that none of these potentials form bound states. A potential 
well as in Eq.~(\ref{sphw}) with $R_N =1$ holds an $S-$wave bound state when 
$\alpha_{J/\psi} > 4.4~{\rm GeV}^{-3}$, a value twice as large as the largest value 
commonly practiced in the literature. The situation is different in the case
of a nucleus. Even when the $J/\psi\,N$ interaction is too weak to bind the two
hadrons, nuclear many-body effects play an important role. Of~particular importance
are the effects of the nuclear mean fields on sub-threshold $DD-$states~\cite{Krein:2017usp,
Tsushima:2011kh,Krein:2010vp,Krein:2013rha}. 

%
\section{Conclusions and perspectives}
\label{sec:concl}

We studied the prospects of using femtoscopy in high-energy proton-proton and heavy-ion 
collisions for learning about the low-momentum $J/\psi$-nucleon interaction. Femtoscopic 
correlation measurements offer the opportunity to access information on low-energy 
hadron-hadron forces inaccessible by other means. Within the QCD multipole 
expansion framework, the forward $J/\psi$-nucleon scattering amplitude is given in terms 
of the $J/\psi$ chromopolarizability and a key matrix element to the origin of mass 
problem, the average chromoelectric gluon distribution in the nucleon, which in turn 
relates to the nucleon mass via the QCD trace anomaly. 

Although we focused our study on the $J/\psi$-nucleon system, femtoscopic
measurements can also be performed for other heavy quarkonia. Our choice 
of the $J/\psi$ was motivated by two main facts: the relatively high $J/\psi$ 
production yields and weak decay rates; and the theoretical information available 
from lattice QCD simulations and phenomenological models. We made use of this 
information to compute $J/\psi\,N$ correlation functions. The available 
information points towards a relatively short-ranged, weakly attractive 
interaction. These features of the interaction, together with the QCD multipole 
expansion framework, lead to a direct relationship between the correlation function 
at small momenta and the average chromoelectric gluon distribution 
in the nucleon when using the Lednicky-Lyuboshits (LL) model. Away from the 
LL model, the link between the chromoelectric gluon distribution in the 
nucleon and the femtoscopic momentum correlation function is less direct. 
Although one would still have access to information on the interaction, 
e.g. on scattering parameters, the theoretical interpretation of the data 
would be more subtle. In any case, the strength of the correlation depends on 
the $J/\psi$ chromoelectric polarizabilty, $\alpha_{J/\psi}$, 
a $J/\psi$ property. This quantity, however, is poorly constrained by experiment. 
If its value turns out to be $\alpha_{J/\psi} > 0.24~{\rm GeV}^{-3}$, our model 
calculations indicate a good likelihood for a femtoscopic extraction of the 
average chromoelectric gluon distribution in the nucleon; or else, one will 
have to find other ways to get it.

We have not addressed experimental issues that can impact the extraction of a 
low-momentum $J/\psi$-nucleon correlation function. As already mentioned, nontrivial 
issues include source form and size, momentum resolution, and non-femtoscopic correlations. 
Notwithstanding these issues, we hope the positive prospects of our theoretical study 
motivate an experimental study as well.

\begin{acknowledgements}
Enlightening and useful discussions with Johann Haidenbauer are gratefully acknowledged.
G.K was partially supported by: Conselho Nacional 
de Desenvolvimento Cient\'{\i}fico e Tecnol\'ogico - CNPq, Grants No. Grant. nos. 
309262/2019-4, 464898/2014-5 (INCT F\'{\i}sica 
Nuclear e Apli\-ca\-\c{c}\~oes), Funda\c{c}\~ao de Amparo \`a 
Pesquisa do Estado de S\~ao Paulo - FAPESP, Grant No. 2013/01907-0. T.C.P was 
supported by a scholarship from Conselho Nacional de Desenvolvimento Cient\'{\i}fico
e Tecnol\'ogico - CNPq.
\end{acknowledgements}

\bibliographystyle{spphys}       


\bibliography{femtoscopy_mass}

\end{document}